\documentclass[12pt]{iopart}
\usepackage{graphicx}
\usepackage{iopams}

\begin{document}

\title{The explicit form of the rate function for semi-Markov processes and its contractions}
\author{Yuki Sughiyama and Testuya J. Kobayashi}
\address{Institute of Industrial Science, The University of Tokyo, 4-6-1, Komaba, Meguro-ku, Tokyo, 153-8505 Japan}
%\ead{yuki.sughiyama@gmail.com}
\vspace{10pt}
\begin{indented}
\item{September 2017}
\end{indented}
\begin{abstract}
We derive the explicit form of the rate function for semi-Markov processes. 
Here, the ``random time change trick" plays an essential role. 
Also, by exploiting the contraction principle of the large deviation theory to the explicit form, we show that the fluctuation theorem (Gallavotti-Cohen Symmetry) holds for semi-Markov cases. 
Furthermore, we elucidate that our rate function is an extension of the Level 2.5 rate function for Markov processes to semi-Markov cases. 
\end{abstract}

%\maketitle

\section{Introduction}
The theory for the large deviation property (LDP) is a significant mathematical tool to describe rare events in a sufficiently large system \cite{01,02,03,04}. 
This theory elucidates a behavior of large fluctuations beyond the variance around the convergence value due to the law of large numbers. 
To be more precise, an exponential decay of rare events caused by expansion of the system size is evaluated by the rate function (the large deviation function). 

In terms of statistical mechanics \cite{05}, the rate function expresses the entropy function, which characterizes thermodynamic properties of many-body systems.
In fact, owing to differentiations of the entropy function (the rate function), we can obtain the heat capacity, the equation of state, and so on. 
Accordingly, one of the main purposes of statistical physics is to calculate an explicit form of the rate function for a given microscopic system. 
In equilibrium statistical mechanics, by using the LDP, we consider the limit of increase in the particle number and volume under a constant density; that is, we calculate a ``spatial" thermodynamic limit.
In particular, we derive the explicit form of the entropy function from a given microscopic Hamiltonian. 
On the other hand, in nonequilibrium situations, we focus on the LDP for a long time average statistics on stochastic processes describing nonequilibrium dynamics \cite{06,07,08}; that is, we consider a ``temporal" thermodynamic limit. 
Especially, a large deviation for the flow (current) characterizing a nonequilibrium state plays an essential role. 
A symmetry of the rate function for the flow is known as the fluctuation theorem (FT) \cite{09,10,11,12}, 
which leads to many recent developments in nonequilibrium physics \cite{13,14,15,16}. 

Explicit forms of the rate functions for nonequilibrium dynamics were derived by various approaches. 
As numerical approaches, recently reported were the biased method using the tilted processes \cite{17,18} 
and the cloning technique employing population dynamics \cite{19,20,21}. 
At the other extreme, as analytic approaches, the study by Donsker and Varadhan is well known. 
In their series of papers \cite{22,23,24,25}, they revealed an explicit form of the rate function for the pair empirical measure, i.e., empirical flow (jump), on discrete-time Markov processes. 
Thus, by the contraction \cite{01,02,04} of the explicit form, we can evaluate the rate function for statistics with respect to the flow (i.e., the heat flow and the entropy production). 
Furthermore, an explicit form of the rate function for continuous-time Markov processes was recently derived in several papers \cite{26,27,28} by using various analytic methods. 
This rate function is known as the {\it Level 2.5} rate function, which describes fluctuations of the empirical occupation and the empirical jump. 

Beyond Markov processes, semi-Markov processes have been studied in various fields \cite{29,30,31,32,33,34,35,36}. 
In contrast to Markov processes, the semi-Markov processes have a memory, because the event (jump) occurrence probability depends on the elapse time after the previous event occurs. 
Owing to this property, the semi-Markov processes play a vital role, when we deal with the enzyme kinetics having memory effect \cite{32}, the age-structured population dynamics such that growth rate depends on age of individuals \cite{33}, and so on. 
Furthermore, it was recently reported that the FT holds even for the semi-Markov cases \cite{34,35}. 
In the same context of the FT, the LDP on semi-Markov processes have also been studied \cite{36}. 
In contrast to the Donsker-Varadhan study and the level 2.5 rate function, however, 
the explicit form of the rate function for semi-Markov processes could not been obtained in the above studies. 

The purpose of this study is to reveal the explicit form of the rate function for semi-Markov processes by using a heuristic approach, which means that our derivation of the rate function is not rigorous, but familiar and accessible to physicists. 
Specifically, we derive the rate function for the empirical jump depending on the waiting (sojourn) time. 
Furthermore, from the explicit form, we rederive the FT for semi-Markov processes by using the contraction principle \cite{01,02,04}. 
Here, the direction time independence (DTI) \cite{20,32,36} of processes plays an important role. 
Finally, by using contraction of the explicit form, we show that our rate function can be reduced to the Level 2.5 rate function on Markov processes. 

This paper is organized as follows. 
In the next section, we briefly introduce semi-Markov processes and their properties. 
In section 3, we derive the explicit form of the rate function for semi-Markov processes by employing the ``random time change trick" \cite{37,38}. 
In section 4, by assuming DTI, we show that the rate function can be decomposed into two parts: the point process and the Markov process parts. 
Furthermore, we indicate that this decomposition leads to the FT on semi-Markov processes. 
In section 5, we consider an age representation of the rate function obtained in section 3, which gives an extension of the Level 2.5 rate function to Semi-Markov cases. Also, we show that, under Markov assumption, our rate function can be reduced to the ordinary Level 2.5 rate function through the contraction principle. 
Finally, we summarize this study in section 6. 

\section{Semi-Markov processes (Markov renewal processes)}
Before working on the main result, we devote this section to an introduction of semi-Markov processes (Markov renewal processes) \cite{32,36} and their properties. 
First, suppose a point process \cite{39} on a time interval $\left[0,t\right],\ \left(T\right):=\left\{T_{i}|1\leq i\leq n_{t}\right\}$: $T_{i}\in[0,\infty)$ denotes the inter-event interval between $i-1$th and $i$th events; 
and $T_{1}$ exceptionally represents the time when the first event occurs. 
Here, $n_{t}$ represents the number of events up to the final time $t$, that is, $n_{t}=\max\left\{n|\sum_{i=1}^{n}T_{i}\leq t\right\}$. 
Also, we assign a state $ X_{i}\in\Omega$ for each inter-event interval $T_{i}$, where $\Omega$ is a finite state space. 
Then, consider a combination of the history (series) of the inter-event intervals and the states, $\left(T,X\right):=\left\{T_{i},X_{i}|1\leq i\leq n_{t}\right\}$. 
In this joint process $\left(T,X\right)$, the events represent jump ($X_{i+1}\neq X_{i}$) or reset ($X_{i+1}=X_{i}$) in the state space $\Omega$, 
and the inter-event interval $T_{i}$ is interpreted as the waiting (sojourn) time in the state $X_{i}$ (see Figure 1). 
\begin{figure}[h]
\begin{center}
\includegraphics*[height=3.5cm]{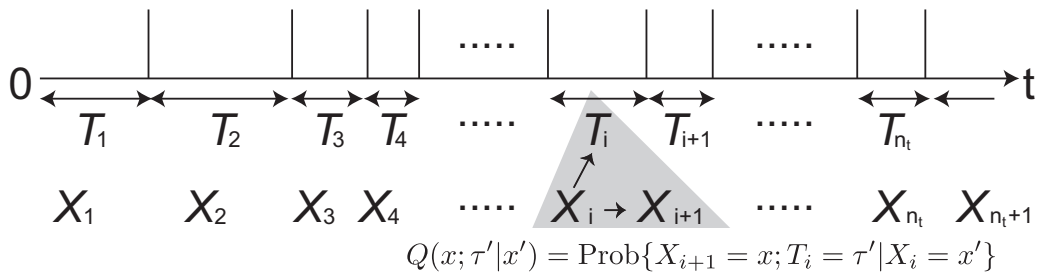}
\caption{Schematic diagram of semi-Markov processes}
\end{center}
\end{figure}
If the updating of the joint process $\left(T,X\right)$ is conditionally independent of the past history given the current state $X_{i}$: 
\begin{eqnarray}
\nonumber\mathrm{Prob}\left\{X_{i+1}=x;T_{i}=\tau^{\prime}|X_{1},...,X_{i};T_{1},...,T_{i-1}\right\}\\
=\mathrm{Prob}\left\{X_{i+1}=x;T_{i}=\tau^{\prime}|X_{i}\right\},
\end{eqnarray}
we call this process a semi-Markov process or a Markov renewal process. 
The generator of this process 
\begin{equation}
Q\left(x;\tau^{\prime}|x^{\prime}\right):=\mathrm{Prob}\left\{X_{i+1}=x;T_{i}=\tau^{\prime}|X_{i}=x^{\prime}\right\},
\end{equation}
is known as semi-Markov kernel \cite{36}. 
Note that this kernel describes the transition probability from $x^{\prime}$ to $x$ after waiting for time $\tau^{\prime}$ in the state $x^{\prime}.$ 
Furthermore, we define a waiting time distribution in the state $x^{\prime}$ as 
\begin{equation}
\displaystyle \pi\left(\tau^{\prime}|x^{\prime}\right):=\sum_{x\in\Omega}Q\left(x;\tau^{\prime}|x^{\prime}\right).\label{pi}
\end{equation}
By using $\pi\left(\tau^{\prime}|x^{\prime}\right)$, the semi-Markov kernel can be decomposed as 
\begin{equation}
Q\left(x;\tau^{\prime}|x^{\prime}\right)=\mathbb{T}\left(x|\tau^{\prime},x^{\prime}\right)\pi\left(\tau^{\prime}|x^{\prime}\right),\label{Qdic}
\end{equation}
where $\mathbb{T}\left(x|\tau^{\prime},x^{\prime}\right)$ is given by the definition of the conditional probability: $\mathbb{T}\left(x|\tau^{\prime},x^{\prime}\right):=Q\left(x;\tau^{\prime}|x^{\prime}\right)/\pi\left(\tau^{\prime}|x^{\prime}\right)$. 
Note that $\mathbb{T}\left(x|\tau^{\prime},x^{\prime}\right)$ expresses the transition probability from $x^{\prime}$ to $x$ under the condition that any event occurs at age $\tau^{\prime}$, 
where the age means the elapsed time after the previous event occurs. 
Also, $\mathbb{T}\left(x|\tau^{\prime},x^{\prime}\right)$ satisfies the property of the transition matrix: $\mathbb{T}\left(x|\tau^{\prime},x\right)=1-\Sigma_{x^{\prime}:x^{\prime}\neq x}\mathbb{T}\left(x^{\prime}|\tau^{\prime},x\right)$. 

Next, to reveal a relationship between $\pi\left(\tau^{\prime}|x^{\prime}\right)$ and the Poisson point process \cite{39}, we rewrite it in terms of an event rate. 
By using $\pi\left(\tau^{\prime}|x^{\prime}\right)$, we can calculate the survival probability up to age $a$ in state $x$, which means the probability that no event occurs up to age $a$: 
\begin{equation}
\displaystyle \Pi\left(a|x\right):=1-\int_{0}^{a}\pi\left(\tau|x\right)d\tau=\int_{a}^{\infty}\pi\left(\tau|x\right)d\tau.\label{Pi}
\end{equation}
Owing to this survival probability, we can represent the semi-Markov kernel as 
\begin{equation}
Q\left(x;\tau^{\prime}|x^{\prime}\right)=r\left(x;\tau^{\prime},x^{\prime}\right)\Pi\left(\tau^{\prime}|x^{\prime}\right),\label{krik1}
\end{equation}
where $r\left(x;\tau^{\prime},x^{\prime}\right)$ represents the probability that a jump event from $x^{\prime}$ to $x$ occurs at age $\tau^{\prime}$, which is known as the hazard function \cite{32}. 
By taking summation with respect to $x$ in (\ref{krik1}) and using (\ref{pi}) and (\ref{Pi}), we have a differential equation: 
\begin{equation}
\displaystyle \frac{d\Pi\left(\tau^{\prime}|x^{\prime}\right)}{d\tau^{\prime}}=-\left\{\sum_{x\in\Omega}r\left(x;\tau^{\prime},x^{\prime}\right)\right\}\Pi\left(\tau^{\prime}|x^{\prime}\right),\label{diff1}
\end{equation}
the solution of which is represented as 
\begin{equation}
\Pi\left(\tau^{\prime}|x^{\prime}\right)=\exp\left[-\int_{0}^{\tau^{\prime}}r\left(a,x^{\prime}\right)da\right].\label{Piex}
\end{equation}
Here, we use the initial condition $\Pi\left(0|x^{\prime}\right)=1$ and $r\left(a,x^{\prime}\right)$ is defined as $r\left(a,x^{\prime}\right):=\Sigma_{x\in\Omega}r\left(x;a,x^{\prime}\right)$, 
which represents the probability that a jump event from $x^{\prime}$ to an arbitrary state occurs at age $a$. 
In this paper, we call $r\left(a,x^{\prime}\right)$ the event rate at age $a$ in the state $x^{\prime}$. 
Accordingly, by differentiating (\ref{Piex}) with respect to $\tau^{\prime}$ and using (\ref{Pi}), 
the waiting time distribution $\pi\left(\tau^{\prime}|x^{\prime}\right)$ can be expressed by the event rate $r\left(a,x^{\prime}\right)$ as 
\begin{equation}
\pi\left(\tau^{\prime}|x^{\prime}\right)=r\left(\tau^{\prime},x^{\prime}\right)\exp\left[-\int_{0}^{\tau^{\prime}}r\left(a,x^{\prime}\right)da\right].\label{pirr}
\end{equation}
Substituting (\ref{pirr}) into (\ref{Qdic}), we can represent the semi-Markov kernel as 
\begin{equation}
Q\left(x;\tau^{\prime}|x^{\prime}\right)=\mathbb{T}\left(x|\tau^{\prime},x^{\prime}\right)r\left(\tau^{\prime},x^{\prime}\right)\exp\left[-\int_{0}^{\tau^{\prime}}r\left(a,x^{\prime}\right)da\right].\label{Qreprr}
\end{equation}
By comparing (\ref{Qreprr}) with (\ref{krik1}) and (\ref{Piex}), we also have $r\left(x;\tau^{\prime},x^{\prime}\right)=\mathbb{T}\left(x|\tau^{\prime},x^{\prime}\right)r\left(\tau^{\prime},x^{\prime}\right)$. 
In addition, if $\mathbb{T}\left(x|\tau^{\prime},x^{\prime}\right)$ is independent of age $\tau^{\prime}$: $\mathbb{T}\left(x|\tau^{\prime},x^{\prime}\right)=\mathbb{T}\left(x|x^{\prime}\right)$, 
we say that the semi-Markov process has {\it direction-time independence} (DTI) \cite{20,32,36}. 

Finally, we mention the connection to continuous-time Markov processes \cite{40,41}. 
If $r\left(a;x^{\prime}\right)$ does not depend on age $a$: $r\left(a,x^{\prime}\right)=r\left(x^{\prime}\right)$ and $\mathbb{T}$ satisfies DTI, then $Q\left(x;\tau^{\prime}|x^{\prime}\right)$ can be written as 
\begin{equation}
Q\left(x;\tau^{\prime}|x^{\prime}\right)=\mathbb{T}\left(x|x^{\prime}\right)r\left(x^{\prime}\right)e^{-r\left(x^{\prime}\right)\tau^{\prime}}.\label{QMarkov}
\end{equation}
This fact represents that the semi-Markov process is reduced to a continuous-time Markov process with transition rate $\omega\left(x|x^{\prime}\right):=\mathbb{T}\left(x|x^{\prime}\right)r\left(x^{\prime}\right)$, 
because any jump event occurs as Poissonian with the event rate $\omega\left(x\right):=\Sigma_{x^{\prime}\in\Omega}\omega\left(x^{\prime}|x\right)=r\left(x\right)$. 

\section{LDP on semi-Markov processes}
We consider the LDP on the semi-Markov process $\left(T,X\right)$. 
In this section, we deal with the following empirical measure for a triplet $\left(x;\tau^{\prime},x^{\prime}\right)$: 
\begin{equation}
j_{e}\displaystyle \left(x;\tau^{\prime},x^{\prime}\right):=\frac{1}{t}\sum_{i=1}^{n_{t}}\delta_{x,X_{i+1}}\delta\left(\tau^{\prime}-T_{i}\right)\delta_{x^{\prime},X_{i}},\label{je3}
\end{equation}
which represents how many times a jump (reset) event from $x^{\prime}$ to $x$ at age $\tau^{\prime}$ occurs in a realization $\left(T,X\right)$. 
Note that this empirical triplet depends on the realization $\left(T,X\right)$, but we abbreviate it from the notation of $j_{e}\left(x;\tau^{\prime},x^{\prime}\right)$ for simplicity. 
Here, we also assume periodic conditions $T_{i+1}=T_{1}$ and $X_{i+1}=X_{1}$. 
This assumption does not restrict generality of the LDP, since the boundary conditions are not effective in the calculation of the empirical measure for $ t\rightarrow\infty$.
Owing to this assumption, the empirical measure satisfies so-called shift-invariant property \cite{01,02,04}: 
\begin{equation}
\displaystyle \sum_{x\in\Omega}\int_{0}^{\infty}d\tau^{\prime}\,j_{e}\left(x;\tau^{\prime},x^{\prime}\right)=\sum_{x\in\Omega}\int_{0}^{\infty}d\tau^{\prime}\,j_{e}\left(x^{\prime};\tau^{\prime},x\right).\label{shiftjg3}
\end{equation}
Also, we introduce a marginal measure: 
\begin{equation}
g_{e}\displaystyle \left(\tau^{\prime},x^{\prime}\right):=\sum_{x\in\Omega}j_{e}\left(x;\tau^{\prime},x^{\prime}\right),
\end{equation}
which quantifies how often any event at age $\tau^{\prime}$ in state $x^{\prime}$ occurs in the realization $\left(T,X\right)$. 
Furthermore, the following property is useful: 
\begin{equation}
\displaystyle \frac{n_{t}}{t}=\sum_{x,x^{\prime}\in\Omega}\int_{0}^{\infty}d\tau^{\prime}\,j_{e}\left(x;\tau^{\prime},x^{\prime}\right)=\sum_{x^{\prime}\in\Omega}\int_{0}^{\infty}d\tau^{\prime}\,g_{e}\left(\tau^{\prime},x^{\prime}\right),\label{ntnorm}
\end{equation}
which represents the number of events per unit time. 
Finally, we shall note a normalization condition of $j_{e}\left(x;\tau^{\prime},x^{\prime}\right)$. 
By using the definition of $j_{e}\left(x;\tau^{\prime},x^{\prime}\right)$, (\ref{je3}), we obtain
\begin{equation}
\displaystyle \sum_{x,x^{\prime}\in\Omega}\int_{0}^{\infty}d\tau^{\prime}\,\tau^{\prime}j_{e}\left(x;\tau^{\prime},x^{\prime}\right)=\frac{1}{t}\sum_{i=1}^{n_{t}}T_{i}.
\end{equation}
If we assume that $t\approx\Sigma_{i=1}^{n_{t}} T_{i}$ for $ t\rightarrow\infty$, which means that we can ignore the elapse time after the final event occurs, 
we find the normalization condition at $ t\rightarrow\infty$ as 
\begin{equation}
1=\displaystyle \sum_{x,x^{\prime}\in\Omega}\int_{0}^{\infty}d\tau^{\prime}\,\tau^{\prime}j_{e}\left(x;\tau^{\prime},x^{\prime}\right)=\sum_{x^{\prime}\in\Omega}\int_{0}^{\infty}d\tau^{\prime}\,\tau^{\prime}g_{e}\left(\tau^{\prime},x^{\prime}\right).\label{norm}
\end{equation} 

In this section, we reveal the explicit form of the rate function for the empirical triplet $j_{e}\left(x;\tau^{\prime},x^{\prime}\right)$, which is defined as 
\begin{equation}
I\displaystyle \left[j\left(x;\tau^{\prime},x^{\prime}\right)\right]:=\lim_{t\rightarrow\infty}-\frac{1}{t}\log \mathrm{Prob}\left\{j_{e}\left(x;\tau^{\prime},x^{\prime}\right)\approx j\left(x;\tau^{\prime},x^{\prime}\right)\right\}.\label{defIj3}
\end{equation}
To calculate the rate function, we employ the following two steps: 
(i) We regard the semi-Markov process as a 2-dimensional Markov process. 
Then, we calculate the rate function for the pair empirical measure \cite{01,02,04} and its contracted one. 
(ii) By using ``random time change trick" \cite{37,38}, we obtain the rate function for the empirical triplet $j_{e}\left(x;\tau^{\prime},x^{\prime}\right)$. 
We describe the above two steps in the next two subsections. 

\subsection{LDP on 2-dimensional Markov processes}
Suppose a 2-dimensional discrete-time Markov process $\left(T,X\right)=\left\{T_{i},X_{i}\right\}$ with a transition probability $M\left(\tau,x|\tau^{\prime},x^{\prime}\right):=\pi\left(\tau|x\right)\mathbb{T}\left(x|\tau^{\prime},x^{\prime}\right)$, 
which is equivalent to the semi-Markov process introduced in the previous section (see Figure 2). 
\begin{figure}[h]
\begin{center}
\includegraphics*[height=3.5cm]{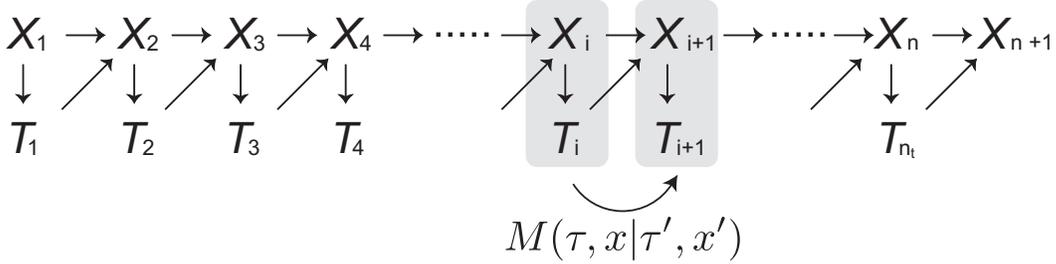}
\caption{Schematic diagram of the 2-dimensional discrete-time Markov process}
\end{center}
\end{figure}
Note that, in terms of this Markov process, $T=\left\{T_{i}\right\}$ is just discrete-time state sequence, i.e., $T_{i}$ does not represent the waiting time. 
Let us consider the 2-dimensional pair empirical measure: 
\begin{equation}
J_{e}\displaystyle \left(\tau,x;\tau^{\prime},x^{\prime}\right):=\frac{1}{n}\sum_{i=1}^{n}\delta\left(\tau-T_{i+1}\right)\delta_{x,X_{i+1}}\delta\left(\tau^{\prime}-T_{i}\right)\delta_{x^{\prime},X_{i}},
\end{equation}
where we again assume periodic conditions $T_{i+1}=T_{1}$ and $X_{i+1}=X_{1}$. 
Here, we note that this pair empirical measure is normalized by the number of events $n$, differently from $j_{e}\left(x;\tau^{\prime},x^{\prime}\right)$ in (\ref{je3}); 
thus $\Sigma_{x,x^{\prime}\in\Omega} \int_{0}^{\infty} d\tau d\tau^{\prime}\,J_{e}\left(\tau,x;\tau^{\prime},x^{\prime}\right)=1$. 
The shift-invariant property is also satisfied: 
\begin{equation}
\displaystyle \sum_{x\in\Omega}\int_{0}^{\infty}d\tau\,J_{e}\left(\tau,x;\tau^{\prime},x^{\prime}\right)=\sum_{x\in\Omega}\int_{0}^{\infty}d\tau\,J_{e}\left(\tau^{\prime},x^{\prime};\tau,x\right)=:G_{e}\left(\tau^{\prime};x^{\prime}\right).\label{shiftJ4}
\end{equation}
If the transition probability $M\left(\tau,x|\tau^{\prime},x^{\prime}\right)$ has ergodicity on 2-dimensional space $\left(\tau,x\right)$, by using Sanov's theorem for Markov processes \cite{01,02,04}, 
the rate function of the pair empirical measure is evaluated as 
\begin{eqnarray}
\displaystyle \nonumber\tilde{I}\left[J\left(\tau,x;\tau^{\prime},x^{\prime}\right)\right]&=&\displaystyle \sum_{x,x^{\prime}\in\Omega}\int_{0}^{\infty}\!\!\!\int_{0}^{\infty}d\tau d\tau^{\prime}\,J\left(\tau,x;\tau^{\prime},x^{\prime}\right)\\
&&\displaystyle \times\log\frac{J\left(\tau,x;\tau^{\prime},x^{\prime}\right)}{M\left(\tau,x|\tau^{\prime},x^{\prime}\right)G\left(\tau^{\prime};x^{\prime}\right)}.\label{IJ4}
\end{eqnarray}
A brief derivation of this rate function is shown in Appendix A, and the well-known rigorous proof is in references \cite{01,02}. 
In this study, we only deal with cases where $M\left(\tau,x|\tau^{\prime},x^{\prime}\right)=\pi\left(\tau|x\right)\mathbb{T}\left(x|\tau^{\prime},x^{\prime}\right)$ satisfies ergodicity (i.e., $\pi$ and $\mathbb{T}$ are restricted). 
By substituting the definition of $M\left(\tau,x|\tau^{\prime},x^{\prime}\right)$ into (\ref{IJ4}), 
we can rewrite the rate function as
\begin{eqnarray}
\displaystyle \nonumber\tilde{I}\left[J\left(\tau,x;\tau^{\prime},x^{\prime}\right)\right]&=&\displaystyle \sum_{x,x^{\prime}\in\Omega}\int_{0}^{\infty}\!\!\!\int_{0}^{\infty}d\tau d\tau^{\prime}\,J\left(\tau,x;\tau^{\prime},x^{\prime}\right)\\
\displaystyle \nonumber&&\times\log\frac{J\left(\tau,x;\tau^{\prime},x^{\prime}\right)}{\pi\left(\tau|x\right)\mathbb{T}\left(x|\tau^{\prime},x^{\prime}\right)G\left(\tau^{\prime};x^{\prime}\right)}\\
\displaystyle \nonumber&=&\displaystyle \sum_{x,x^{\prime}\in\Omega}\int_{0}^{\infty}\!\!\!\int_{0}^{\infty}d\tau d\tau^{\prime}\,J\left(\tau,x;\tau^{\prime},x^{\prime}\right)\\
\displaystyle \nonumber&&\times\log\frac{J\left(\tau,x;\tau^{\prime},x^{\prime}\right)}{\mathbb{T}\left(x|\tau^{\prime},x^{\prime}\right)\pi\left(\tau^{\prime}|x^{\prime}\right)G\left(\tau^{\prime};x^{\prime}\right)}\\
\displaystyle \nonumber&=&\displaystyle \sum_{x,x^{\prime}\in\Omega}\int_{0}^{\infty}\!\!\!\int_{0}^{\infty}d\tau d\tau^{\prime}\,J\left(\tau,x;\tau^{\prime},x^{\prime}\right)\\
&&\displaystyle \times\log\frac{J\left(\tau,x;\tau^{\prime},x^{\prime}\right)}{Q\left(x;\tau^{\prime}|x^{\prime}\right)G\left(\tau^{\prime};x^{\prime}\right)},\label{IJ4m}
\end{eqnarray}
where we use the representation of the semi-Markov kernel (\ref{Qdic}). 
In the second line in (\ref{IJ4m}), we also use the equality, 
\begin{eqnarray}
\displaystyle \nonumber\sum_{x,x^{\prime}\in\Omega}\int_{0}^{\infty}d\tau d\tau^{\prime}\,J\left(\tau,x;\tau^{\prime},x^{\prime}\right)\log\frac{1}{\pi\left(\tau|x\right)}\\
=\displaystyle \sum_{x,x^{\prime}\in\Omega}\int_{0}^{\infty}d\tau d\tau^{\prime}\,J\left(\tau,x;\tau^{\prime},x^{\prime}\right)\log\frac{1}{\pi\left(\tau^{\prime}|x^{\prime}\right)},\label{kikikikit}
\end{eqnarray}
which is derived by the shift-invariant property (\ref{shiftJ4}). 

Next, we consider the following useful decomposition of the joint probability $J\left(\tau,x;\tau^{\prime},x^{\prime}\right)$ for contraction: 
\begin{equation}
J\left(\tau,x;\tau^{\prime},x^{\prime}\right)=J\left(\tau|x;\tau^{\prime},x^{\prime}\right)J\left(x;\tau^{\prime},x^{\prime}\right),\label{Jdic}
\end{equation}
where $J\left(x;\tau^{\prime},x^{\prime}\right)$ is defined as $J\left(x;\tau^{\prime},x^{\prime}\right)= \int_{0}^{\infty} d\tau\,J\left(\tau,x;\tau^{\prime},x^{\prime}\right)$ 
and the conditional probability $J\left(\tau|x;\tau^{\prime},x^{\prime}\right)$ is given by the definition of the conditional probability: $J\left(\tau|x;\tau^{\prime},x^{\prime}\right):=J\left(\tau,x;\tau^{\prime},x^{\prime}\right)/J\left(x;\tau^{\prime},x^{\prime}\right)$. 
Also, we decompose $G\left(\tau^{\prime};x^{\prime}\right)$ as 
\begin{equation}
G\left(\tau^{\prime};x^{\prime}\right)=G\left(\tau^{\prime}|x^{\prime}\right)G\left(x^{\prime}\right),\label{Gdic}
\end{equation}
where we again use $G\left(\tau^{\prime}|x^{\prime}\right):=G\left(\tau^{\prime};x^{\prime}\right)/G\left(x^{\prime}\right)$. 
By substituting (\ref{Jdic}) and (\ref{Gdic}) into (\ref{IJ4m}), we have
\begin{eqnarray}
\displaystyle \nonumber\tilde{I}\left[J\left(\tau,x;\tau^{\prime},x^{\prime}\right)\right]&=&\displaystyle \sum_{x,x^{\prime}\in\Omega}\int_{0}^{\infty}d\tau^{\prime}\,J\left(x;\tau^{\prime},x^{\prime}\right)\log\frac{J\left(x;\tau^{\prime},x^{\prime}\right)}{Q\left(x;\tau^{\prime}|x^{\prime}\right)G\left(x^{\prime}\right)}\\
&&+\displaystyle \sum_{x,x^{\prime}\in\Omega}\int_{0}^{\infty}d\tau d\tau^{\prime}\,J\left(\tau,x;\tau^{\prime},x^{\prime}\right)\log\frac{J\left(\tau|x;\tau^{\prime},x^{\prime}\right)}{G\left(\tau|x\right)},\label{IJ4mm}
\end{eqnarray}
where, in the second line, we change the argument of $G$ from $G\left(\tau^{\prime}|x^{\prime}\right)$ to $G\left(\tau|x\right)$ by using the shift-invariant property of $J\left(\tau,x;\tau^{\prime},x^{\prime}\right)$ as in (\ref{kikikikit}). 
By employing the above preparation, we calculate the rate function of the empirical measure for the triplet $\left(x;\tau^{\prime},x^{\prime}\right)$, 
\begin{equation}
J_{e}\displaystyle \left(x;\tau^{\prime},x^{\prime}\right):=\frac{1}{n}\sum_{i=1}^{n}\delta_{x,X_{i+1}}\delta\left(\tau^{\prime}-T_{i}\right)\delta_{x^{\prime},X_{i}}=\int_{0}^{\infty}d\tau\,J_{e}\left(\tau,x;\tau^{\prime},x^{\prime}\right),\label{Je3}
\end{equation}
which measures how often a triplet $\left(x;\tau^{\prime},x^{\prime}\right)$ appears in a realization $\left(T,X\right)$. 
Owing to the contraction principle of the LDP \cite{01,02,04}, the rate function is evaluated as 
\begin{eqnarray}
\displaystyle \nonumber\tilde{I}\left[J\left(x;\tau^{\prime},x^{\prime}\right)\right]=\min_{J\left(\tau,x;\tau^{\prime},x^{\prime}\right)}\biggl\{&&I\displaystyle \left[J\left(\tau,x;\tau^{\prime},x^{\prime}\right)\right]|\\
&&\displaystyle \int_{0}^{\infty}d\tau\,J\left(\tau,x;\tau^{\prime},x^{\prime}\right)=J\left(x;\tau^{\prime},x^{\prime}\right)\biggr\}.
\end{eqnarray}
Since $J\left(x;\tau^{\prime},x^{\prime}\right)$ is fixed in the minimization, we can only change the conditional probability $J\left(\tau|x;\tau^{\prime},x^{\prime}\right)$ for the minimization, see (\ref{Jdic}). 
Thus, if we can choose $J\left(\tau|x;\tau^{\prime},x^{\prime}\right)$ as $G\left(\tau|x\right)$, the second term of (\ref{IJ4mm}) can be eliminated, and we have $\tilde{I}\left[J\left(x;\tau^{\prime},x^{\prime}\right)\right]$ as 
\begin{equation}
\displaystyle \tilde{I}\left[J\left(x;\tau^{\prime},x^{\prime}\right)\right]=\sum_{x,x^{\prime}\in\Omega}\int_{0}^{\infty}d\tau^{\prime}\,J\left(x;\tau^{\prime},x^{\prime}\right)\log\frac{J\left(x;\tau^{\prime},x^{\prime}\right)}{Q\left(x;\tau^{\prime}|x^{\prime}\right)G\left(x^{\prime}\right)},\label{IJ3}
\end{equation}
where we note that the following shift-invariant property is satisfied:
\begin{equation}
\displaystyle \sum_{x\in\Omega}\int_{0}^{\infty}d\tau^{\prime}\,J\left(x;\tau^{\prime},x^{\prime}\right)=\sum_{x\in\Omega}\int_{0}^{\infty}d\tau^{\prime}\,J\left(x^{\prime};\tau^{\prime},x\right)=G\left(x^{\prime}\right),
\end{equation}
where we use (\ref{shiftJ4}).
By using the remaining part of this subsection, we prove the reason why we can choose $J\left(\tau|x;\tau^{\prime},x^{\prime}\right)$ as $G\left(\tau|x\right)$. 
Due to the shift-invariant property (\ref{shiftJ4}), the conditional probability $J\left(\tau|x;\tau^{\prime},x^{\prime}\right)$ must satisfy 
\begin{eqnarray}
\displaystyle \nonumber\sum_{x\in\Omega}\int_{0}^{\infty}d\tau\,J\left(\tau|x;\tau^{\prime},x^{\prime}\right)J\left(x;\tau^{\prime},x^{\prime}\right)=\sum_{x\in\Omega}\int_{0}^{\infty}d\tau\,J\left(\tau^{\prime}|x^{\prime};\tau,x\right)J\left(x^{\prime};\tau,x\right).\\
\end{eqnarray}
In order to choose $J\left(\tau|x;\tau^{\prime},x^{\prime}\right)=G\left(\tau|x\right)$, we have to verify that the following equality holds: 
\begin{equation}
\displaystyle \sum_{x\in\Omega}\int_{0}^{\infty}d\tau\,G\left(\tau|x\right)J\left(x;\tau^{\prime},x^{\prime}\right)=\sum_{x\in\Omega}\int_{0}^{\infty}d\tau\,G\left(\tau^{\prime}|x^{\prime}\right)J\left(x^{\prime};\tau,x\right).
\end{equation}
The left hand side is calculated as $\Sigma_{x\in\Omega}J\left(x;\tau^{\prime},x^{\prime}\right)=G\left(\tau^{\prime},x^{\prime}\right)$ by the definition of $G\left(\tau^{\prime},x^{\prime}\right)$, (\ref{shiftJ4}); 
on the other hand, the right hand side is evaluated as $G\left(\tau^{\prime}|x^{\prime}\right)G\left(x^{\prime}\right)=G\left(\tau^{\prime},x^{\prime}\right)$. 
Accordingly, we can choose $J\left(\tau|x;\tau^{\prime},x^{\prime}\right)=G\left(\tau|x\right)$, and therefore the rate function of the empirical triplet $\left(x;\tau^{\prime},x^{\prime}\right)$ can be represented by (\ref{IJ3}). 

\subsection{Random time change}
As shown in the previous subsection, the rate function of $J_{e}\left(x;\tau^{\prime},x^{\prime}\right)$ is given by (\ref{IJ3}). 
However, since $J_{e}\left(x;\tau^{\prime},x^{\prime}\right)$ is normalized by the number of events $n$, we need to change $J_{e}\left(x;\tau^{\prime},x^{\prime}\right)$ to $j_{e}\left(x;\tau^{\prime},x^{\prime}\right)$, 
the later of which is normalized by time $t$, see (\ref{je3}). 
In this subsection, we consider a scaling of $\tilde{I}\left[J\left(x;\tau^{\prime},x^{\prime}\right)\right]$. 

We begin with the definition of the time-normalized rate function (\ref{defIj3}). 
By substituting (\ref{je3}) into (\ref{defIj3}), we get
\begin{eqnarray}
\displaystyle \nonumber I\left[j\left(x;\tau^{\prime},x^{\prime}\right)\right]:=\lim_{t\rightarrow\infty}-\frac{1}{t}\log \mathrm{Prob}\left\{j_{e}\left(x;\tau^{\prime},x^{\prime}\right)\approx j\left(x;\tau^{\prime},x^{\prime}\right)\right\}\\
=\displaystyle \lim_{t\rightarrow\infty}-\frac{1}{t}\log \mathrm{Prob}\biggl\{\frac{1}{t}\sum_{i=1}^{n_{t}}\delta_{x,X_{i+1}}\delta\left(\tau^{\prime}-T_{i}\right)\delta_{x^{\prime},X_{i}}\approx j\left(x;\tau^{\prime},x^{\prime}\right)\biggr\}
\end{eqnarray}
By using (\ref{ntnorm}), we have 
\begin{eqnarray}
\displaystyle \nonumber I\left[j\left(x;\tau^{\prime},x^{\prime}\right)\right]=\lim_{t\rightarrow\infty}-\frac{1}{t}\log&&\displaystyle \mathrm{Prob}\biggl\{\frac{1}{t}\sum_{i=1}^{t\sum_{x,x^{\prime}\in\Omega}\int_{0}^{\infty}d\tau^{\prime}\,\tau^{\prime}j\left(x;\tau^{\prime},x^{\prime}\right)}\\
&&\delta_{x,X_{i+1}}\delta\left(\tau^{\prime}-T_{i}\right)\delta_{x^{\prime},X_{i}}\approx j\left(x;\tau^{\prime},x^{\prime}\right)\biggr\}.
\end{eqnarray}
Dividing both sides of the equation in $\mathrm{Prob}\left\{\cdot\right\}$ by $\Sigma_{x,x^{\prime}\in\Omega} \int_{0}^{\infty} d\tau^{\prime}\,j\left(x;\tau^{\prime},x^{\prime}\right)$, we reach 
\begin{eqnarray}
\displaystyle \nonumber I\left[j\left(x;\tau^{\prime},x^{\prime}\right)\right]=\lim_{t\rightarrow\infty}-\frac{1}{t}&&\displaystyle \frac{\sum_{x,x^{\prime}\in\Omega}\int_{0}^{\infty}d\tau^{\prime}\,\tau^{\prime}j\left(x;\tau^{\prime},x^{\prime}\right)}{\sum_{x,x^{\prime}\in\Omega}\int_{0}^{\infty}d\tau^{\prime}\,\tau^{\prime}j\left(x;\tau^{\prime},x^{\prime}\right)}\\
\displaystyle \nonumber&&\times\log \mathrm{Prob}\biggl\{\frac{1}{t\sum_{x,x^{\prime}\in\Omega}\int_{0}^{\infty}d\tau^{\prime}\,\tau^{\prime}j\left(x;\tau^{\prime},x^{\prime}\right)}\\
\displaystyle \nonumber&&\times\sum_{i=1}^{t\sum_{x,x^{\prime}\in\Omega}\int_{0}^{\infty}d\tau^{\prime}\,\tau^{\prime}j\left(x;\tau^{\prime},x^{\prime}\right)}\delta_{x,X_{i+1}}\delta\left(\tau^{\prime}-T_{i}\right)\delta_{x^{\prime},X_{i}}\\
&&\displaystyle \approx\frac{j\left(x;\tau^{\prime},x^{\prime}\right)}{\sum_{x,x^{\prime}\in\Omega}\int_{0}^{\infty}d\tau^{\prime}\,j\left(x;\tau^{\prime},x^{\prime}\right)}\biggr\},
\end{eqnarray}
where we insert $\Sigma_{x,x^{\prime}\in\Omega} \int_{0}^{\infty} d\tau^{\prime}\,j\left(x;\tau^{\prime},x^{\prime}\right)/\Sigma_{x,x^{\prime}\in\Omega} \int_{0}^{\infty} d\tau^{\prime}\,j\left(x;\tau^{\prime},x^{\prime}\right)=1$ before the symbol ``$\log$". 
By defining $n:=t\Sigma_{x,x^{\prime}\in\Omega} \int_{0}^{\infty} d\tau^{\prime}\,j\left(x;\tau^{\prime},x^{\prime}\right)$ and changing the limit $ t\rightarrow\infty$ to $ n\rightarrow\infty$, we have 
\begin{eqnarray}
\displaystyle \nonumber I\left[j\left(x;\tau^{\prime},x^{\prime}\right)\right]&=&\displaystyle \left[\sum_{x,x^{\prime}\in\Omega}\int_{0}^{\infty}d\tau^{\prime}\,j\left(x;\tau^{\prime},x^{\prime}\right)\right]\lim_{n\rightarrow\infty}-\frac{1}{n}\log \mathrm{Prob}\biggl\{\frac{1}{n}\sum_{i=1}^{n}\\
\displaystyle \nonumber&&\delta_{x,X_{i+1}}\delta\left(\tau^{\prime}-T_{i}\right)\delta_{x^{\prime},X_{i}}\approx\frac{j\left(x;\tau^{\prime},x^{\prime}\right)}{\sum_{x,x^{\prime}\in\Omega}\int_{0}^{\infty}d\tau^{\prime}\,j\left(x;\tau^{\prime},x^{\prime}\right)}\biggr\}.\\
\end{eqnarray}
Furthermore, from (\ref{Je3}), we obtain 
\begin{eqnarray}
\displaystyle \nonumber I\left[j\left(x;\tau^{\prime},x^{\prime}\right)\right]&=&\displaystyle \left[\sum_{x,x^{\prime}\in\Omega}\int_{0}^{\infty}d\tau^{\prime}\,j\left(x;\tau^{\prime},x^{\prime}\right)\right]\lim_{t\rightarrow\infty}-\frac{1}{n_{t}}\log\\
&&\displaystyle \mathrm{Prob}\biggl\{J_{e}\left(x;\tau^{\prime},x^{\prime}\right)\approx\frac{j\left(x;\tau^{\prime},x^{\prime}\right)}{\sum_{x,x^{\prime}\in\Omega}\int_{0}^{\infty}d\tau^{\prime}\,j\left(x;\tau^{\prime},x^{\prime}\right)}\biggr\}.
\end{eqnarray}
Finally, taking the definition of the rate function $\tilde{I}\left[J\left(x;\tau^{\prime},x^{\prime}\right)\right]$ into account, 
we find the scaled equation: 
\begin{eqnarray}
\displaystyle \nonumber I\left[j\left(x;\tau^{\prime},x^{\prime}\right)\right]&=&\displaystyle \sum_{x,x^{\prime}\in\Omega}\int_{0}^{\infty}d\tau^{\prime}\,j\left(x;\tau^{\prime},x^{\prime}\right)\\
&&\times\tilde{I}\left[\frac{j\left(x;\tau^{\prime},x^{\prime}\right)}{\sum_{x,x^{\prime}\in\Omega}\int_{0}^{\infty}d\tau^{\prime}\,j\left(x;\tau^{\prime},x^{\prime}\right)}\right].\label{scaling}
\end{eqnarray}
Accordingly, by substituting the explicit form of the rate function $\tilde{I}$, (\ref{IJ3}), into (\ref{scaling}), 
we obtain the rate function of the empirical triplet $j_{e}\left(x;\tau^{\prime},x^{\prime}\right)$ as 
\begin{equation}
I\displaystyle \left[j\left(x;\tau^{\prime},x^{\prime}\right)\right]=\sum_{x,x^{\prime}\in\Omega}\int_{0}^{\infty}d\tau^{\prime}\,j\left(x;\tau^{\prime},x^{\prime}\right)\log\frac{j\left(x;\tau^{\prime},x^{\prime}\right)}{Q\left(x;\tau^{\prime}|x^{\prime}\right)g\left(x^{\prime}\right)},\label{Ij3}
\end{equation}
where $g\left(x^{\prime}\right)$ is defined by the shift-invariant property (\ref{shiftjg3}) as 
\begin{eqnarray}
\displaystyle \nonumber g\left(x^{\prime}\right)&:=&\displaystyle \int_{0}^{\infty}d\tau^{\prime}\,g\left(\tau^{\prime},x^{\prime}\right)\\
&=&\displaystyle \sum_{x\in\Omega}\int_{0}^{\infty}d\tau^{\prime}\,j\left(x;\tau^{\prime},x^{\prime}\right)=\sum_{x\in\Omega}\int_{0}^{\infty}d\tau^{\prime}\,j\left(x^{\prime};\tau^{\prime},x\right).\label{shiftggjj}
\end{eqnarray}
This explicit form of the rate function (\ref{Ij3}) constitutes the foundation of our study. 
In the remaining part of this paper, we will derive various important rate functions by employing contraction for this explicit form. 

The scaling trick used in this subsection is known as ``random time change". 
Although we only give a brief procedure of the random time change in this paper, its rigorous proof is shown in references \cite{37,38}. 

\section{DTI semi-Markov processes and fluctuation theorem}
In this section, we consider LDP on a semi-Markov process with DTI, $\mathbb{T}\left(x|\tau^{\prime},x^{\prime}\right)=\mathbb{T}\left(x|x^{\prime}\right)$. 
In this case, by employing the contraction principle, we can obtain an explicit form of the rate function for the following two empirical measures:
\begin{eqnarray}
g_{e}\displaystyle \left(\tau^{\prime},x^{\prime}\right)&=&\displaystyle \sum_{x\in\Omega}j_{e}\left(x;\tau^{\prime},x^{\prime}\right)=\frac{1}{t}\sum_{i=1}^{n_{t}}\delta\left(\tau^{\prime}-T_{i}\right)\delta_{x^{\prime},X_{i}},\label{gdef2}\\
c_{e}\displaystyle \left(x;x^{\prime}\right)&:=&\displaystyle \int_{0}^{\infty}d\tau^{\prime}\,j_{e}\left(x;\tau^{\prime},x^{\prime}\right)=\frac{1}{t}\sum_{i=1}^{n_{t}}\delta_{x,X_{i+1}}\delta_{x^{\prime},X_{i}},\label{c2xx}
\end{eqnarray}
where $c_{e}\left(x;x^{\prime}\right)$ measures how often a jump (reset) from $x^{\prime}$ to $x$ occurs in the realization $\left(T,X\right)$ and satisfies the shift-invariant property: 
\begin{equation}
\displaystyle \sum_{x\in\Omega}c_{e}\left(x;x^{\prime}\right)=\sum_{x\in\Omega}c_{e}\left(x^{\prime};x\right)=g_{e}\left(x^{\prime}\right).\label{shiftccg}
\end{equation}
From these empirical measures, we can see that the rate function is composed of two parts: rate functions of point processes and Markov processes. 
Furthermore, by using the explicit form obtained, we show that the fluctuation theorem (Gallavotti-Cohen Symmetry) \cite{09,10} holds even for DTI semi-Markov cases. 

\subsection{Rate function for DTI semi-Markov processes }
We start with the rate function (\ref{Ij3}). 
Since we now consider the DTI case, we substitute $Q\left(x;\tau^{\prime}|x^{\prime}\right)=\mathbb{T}\left(x|x^{\prime}\right)\pi\left(\tau^{\prime}|x^{\prime}\right)$ into (\ref{Ij3}); then we have
\begin{equation}
I\displaystyle \left[j\left(x;\tau^{\prime},x^{\prime}\right)\right]=\sum_{x,x^{\prime}\in\Omega}\int_{0}^{\infty}d\tau^{\prime}\,j\left(x;\tau^{\prime},x^{\prime}\right)\log\frac{j\left(x;\tau^{\prime},x^{\prime}\right)}{\mathbb{T}\left(x|x^{\prime}\right)\pi\left(\tau^{\prime}|x^{\prime}\right)g\left(x^{\prime}\right)}.\label{Ij3DTI}
\end{equation}
To prepare for the following calculations, we introduce decompositions:
\begin{eqnarray}
j\left(x;\tau^{\prime},x^{\prime}\right)&=&j\left(\tau^{\prime}|x;x^{\prime}\right)c\left(x;x^{\prime}\right),\label{dicj3}\\
g\left(\tau^{\prime},x^{\prime}\right)&=&g\left(\tau^{\prime}|x^{\prime}\right)g\left(x^{\prime}\right),\label{dicg2}
\end{eqnarray}
where $j\left(\tau^{\prime}|x;x^{\prime}\right)$ and $g\left(\tau^{\prime}|x^{\prime}\right)$ are conditional measures. 
Substituting (\ref{dicj3}) and (\ref{dicg2}) into (\ref{Ij3DTI}), we get 
\begin{eqnarray}
\displaystyle \nonumber I\left[j\left(x;\tau^{\prime},x^{\prime}\right)\right]&=&\displaystyle \sum_{x^{\prime}\in\Omega}\int_{0}^{\infty}d\tau^{\prime}\,g\left(\tau^{\prime},x^{\prime}\right)\log\frac{g\left(\tau^{\prime},x^{\prime}\right)}{\pi\left(\tau^{\prime}|x^{\prime}\right)g\left(x^{\prime}\right)}\\
\displaystyle \nonumber&&+\sum_{x,x^{\prime}\in\Omega}c\left(x;x^{\prime}\right)\log\frac{c\left(x;x^{\prime}\right)}{\mathbb{T}\left(x|x^{\prime}\right)g\left(x^{\prime}\right)}\\
&&+\displaystyle \sum_{x,x^{\prime}\in\Omega}\int_{0}^{\infty}d\tau^{\prime}\,j\left(x;\tau^{\prime},x^{\prime}\right)\log\frac{j\left(\tau^{\prime}|x;x^{\prime}\right)}{g\left(\tau^{\prime}|x^{\prime}\right)}.
\end{eqnarray}
Now, the rate function for the empirical measures $g_{e}\left(\tau^{\prime},x^{\prime}\right)$ and $c_{e}\left(x;x^{\prime}\right)$ is given by the contraction principle as 
\begin{eqnarray}
\displaystyle \nonumber I\left[g\left(\tau^{\prime},x^{\prime}\right),c\left(x;x^{\prime}\right)\right]=\min_{j\left(x;\tau^{\prime},x^{\prime}\right)}\biggl\{&&I\displaystyle \left[j\left(x;\tau^{\prime},x^{\prime}\right)\right]|\sum_{x\in\Omega}j\left(x;\tau^{\prime},x^{\prime}\right)=g\left(\tau^{\prime},x^{\prime}\right),\\
&&\displaystyle \int_{0}^{\infty}d\tau^{\prime}\,j\left(x;\tau^{\prime},x^{\prime}\right)=c\left(x;x^{\prime}\right)\biggr\}.
\end{eqnarray}
Since $g\left(\tau,x\right)$ and $c\left(x;x^{\prime}\right)$ are fixed in the minimization, we can only sweep the conditional measure $j\left(\tau^{\prime}|x;x^{\prime}\right)$ (see (\ref{dicj3})) 
under a constraint that the following equation holds: 
\begin{equation}
\displaystyle \sum_{x\in\Omega}j\left(\tau^{\prime}|x;x^{\prime}\right)c\left(x;x^{\prime}\right)=g\left(\tau^{\prime},x^{\prime}\right).\label{xxx1}
\end{equation}
By choosing $j\left(\tau^{\prime}|x;x^{\prime}\right)$ as $j\left(\tau^{\prime}|x;x^{\prime}\right)=g\left(\tau^{\prime}|x^{\prime}\right)$, we obtain the rate function for $g\left(\tau,x\right)$ and $c\left(x;x^{\prime}\right)$ as 
\begin{eqnarray}
\displaystyle \nonumber I\left[g\left(\tau,x\right),c\left(x;x^{\prime}\right)\right]&=&\displaystyle \sum_{x\in\Omega}\int_{0}^{\infty}d\tau\,g\left(\tau,x\right)\log\frac{g\left(\tau,x\right)}{\pi\left(\tau|x\right)g\left(x\right)}\\
&&+\displaystyle \sum_{x,x^{\prime}\in\Omega}c\left(x;x^{\prime}\right)\log\frac{c\left(x;x^{\prime}\right)}{\mathbb{T}\left(x|x^{\prime}\right)g\left(x^{\prime}\right)},\label{Igc}
\end{eqnarray}
where we change the summation index and the integration variable in the first term from $\left(\tau^{\prime},x^{\prime}\right)$ to $\left(\tau,x\right)$. 
Here, we note that the choice $j\left(\tau^{\prime}|x;x^{\prime}\right)=g\left(\tau^{\prime}|x^{\prime}\right)$ satisfies (\ref{xxx1}) because we can have
\begin{equation}
\displaystyle \sum_{x\in\Omega}g\left(\tau^{\prime}|x^{\prime}\right)c\left(x;x^{\prime}\right)=g\left(\tau^{\prime}|x^{\prime}\right)g\left(x^{\prime}\right)=g\left(\tau^{\prime},x^{\prime}\right).
\end{equation} 
The rate function (\ref{Igc}) is composed of two terms.
The first term describes the rate function on point processes, which determines the inter-event interval. 
An explanation of point processes and their LDP is shown in Appendix B. 
On the other hand, the second term represents the rate function for Markov jump processes, 
which is the same form as the rate function for pair empirical measure on discreet-time Markov processes, (\ref{aadd1}), in Appendix A. 
Owing to this explicit form, we can find the fluctuation theorem for DTI semi-Markov processes as in the next subsection. 

\subsection{Fluctuation theorem (Gallavotti-Cohen Symmetry)}
Various significant developments in the statistical physics have recently been brought by the fluctuation theorem (FT) \cite{09,10,11,12,13,14,15,16}, 
which describes the time reversal symmetry of the entropy production. 
(To be more precise, it originally expresses the symmetry of the entropy flow (current); 
however, in nonequilibrium stationary situation, the entropy flow is equivalent to the entropy production.) 
Especially, in terms of LDP, the FT appears as a symmetry of the rate function for the entropy production, 
which is called Gallavotti-Cohen Symmetry (GCS) \cite{09,10}. 
Although many studies concerned the FT on Markov processes, 
some recent studies elucidate that FT can be extended to the case of DTI semi-Markov processes \cite{32,34,35,36}. 
In this subsection, by using the explicit form (\ref{Igc}), we show that the GCS holds on DTI semi-Markov processes; 
which was originally proved by Maes {\it et al}. \cite{36}, by using a different approach from ours. 

According to several studies \cite{32,36} treating the FT on semi-Markov processes, under the DTI assumption, 
the entropy production (flow) associated with a jump from $x^{\prime}$ to $x$ is represented as 
\begin{equation}
\displaystyle \Sigma\left(x;x^{\prime}\right):=\log\frac{\mathbb{T}\left(x|x^{\prime}\right)\theta\left(x^{\prime}\right)}{\mathbb{T}\left(x^{\prime}|x\right)\theta\left(x\right)},\label{LDB}
\end{equation}
where $\theta\left(x\right)$ describes the effective escape rate from state $x$, which is defined by 
\begin{equation}
\displaystyle \frac{1}{\theta\left(x\right)}:=\int_{0}^{\infty}d\tau\,\tau\pi\left(\tau|x\right)=\int_{0}^{\infty}d\tau\,\Pi\left(\tau|x\right).\label{thpi}
\end{equation}
Here, we use 
\begin{equation}
\displaystyle \int_{0}^{\infty}d\tau\,\Pi\left(\tau|x\right)=\int_{0}^{\infty}d\tau\int_{\tau}^{\infty}dt\,\pi\left(t|x\right)=\int_{0}^{\infty}dt\,t\pi\left(t|x\right).
\end{equation}
Equations (\ref{LDB}) and (\ref{thpi}) respectively indicate extensions of the detailed fluctuation theorem (local detailed balance) and the escape rate to semi-Markov cases. 
If we assume the Markov condition $r\left(a,x\right)=r\left(x\right)$, 
the integration in (\ref{thpi}) is calculated as 
\begin{equation}
\displaystyle \int_{0}^{\infty}d\tau\,\Pi\left(\tau|x\right)=\int_{0}^{\infty}d\tau\,e^{-r\left(x\right)\tau}=\frac{1}{r\left(x\right)},
\end{equation}
where we use (\ref{Piex}). 
Therefore, $\theta\left(x\right)$ can be reduced to the ordinary escape rate of Markov processes, $\theta\left(x\right)=r\left(x\right)$ ($=\omega\left(x\right)$). 
Furthermore, recalling that $\mathbb{T}\left(x|x^{\prime}\right)r\left(x^{\prime}\right)$ expresses the transition rate of Markov processes, $\omega\left(x|x^{\prime}\right):=\mathbb{T}\left(x|x^{\prime}\right)r\left(x^{\prime}\right)$, 
we find that equation (\ref{LDB}) is reduced to the well-known detailed fluctuation theorem on Markov processes. 

Consider the time-averaged entropy production rate on a sufficient long path $\left(T,X\right)$:
\begin{equation}
\displaystyle \sigma_{e}:=\lim_{t\rightarrow\infty}\frac{1}{t}\sum_{i=1}^{n_{t}}\Sigma\left(X_{i+1},X_{i}\right)=\sum_{x,x^{\prime}\in\Omega}c_{e}\left(x,x^{\prime}\right)\Sigma\left(x;x^{\prime}\right).
\end{equation}
Since $c_{e}\left(x,x^{\prime}\right)$ has the shift-invariant property (\ref{shiftccg}), we get 
\begin{equation}
\displaystyle \sigma_{e}=\sum_{x,x^{\prime}\in\Omega}c_{e}\left(x,x^{\prime}\right)\log\frac{\mathbb{T}\left(x|x^{\prime}\right)\theta\left(x^{\prime}\right)}{\mathbb{T}\left(x^{\prime}|x\right)\theta\left(x\right)}=\sum_{x,x^{\prime}\in\Omega}c_{e}\left(x,x^{\prime}\right)\log\frac{\mathbb{T}\left(x|x^{\prime}\right)}{\mathbb{T}\left(x^{\prime}|x\right)}
\end{equation}
We now investigate the symmetry of the rate function for $\sigma_{e}$. 
To do that, we firstly elucidate a relationship between $I\left[g\left(\tau,x\right),c\left(x;x^{\prime}\right)\right]$ and $I\left[g\left(\tau,x\right),c\left(x^{\prime};x\right)\right]$. 
For the notational simplicity, here we write the transpose matrix of $c\left(x;x^{\prime}\right)$ as $c\left(x^{\prime};x\right)$. 
By using (\ref{Igc}), we have
\begin{eqnarray}
\displaystyle \nonumber I\left[g\left(\tau,x\right),c\left(x^{\prime};x\right)\right]&=&\displaystyle \sum_{x\in\Omega}\int_{0}^{\infty}d\tau\,g\left(\tau,x\right)\log\frac{g\left(\tau,x\right)}{\pi\left(\tau|x\right)g\left(x\right)}\\
\displaystyle \nonumber&&+\sum_{x,x^{\prime}\in\Omega}c\left(x^{\prime};x\right)\log\frac{c\left(x^{\prime};x\right)}{\mathbb{T}\left(x|x^{\prime}\right)g\left(x^{\prime}\right)}\\
\displaystyle \nonumber&=&\displaystyle \sum_{x\in\Omega}\int_{0}^{\infty}d\tau\,g\left(\tau,x\right)\log\frac{g\left(\tau,x\right)}{\pi\left(\tau|x\right)g\left(x\right)}\\
\displaystyle \nonumber&&+\sum_{x,x^{\prime}\in\Omega}c\left(x;x^{\prime}\right)\log\frac{c\left(x;x^{\prime}\right)}{\mathbb{T}\left(x|x^{\prime}\right)g\left(x^{\prime}\right)}\\
\displaystyle \nonumber&&+\sum_{x,x^{\prime}\in\Omega}c\left(x,x^{\prime}\right)\log\frac{\mathbb{T}\left(x|x^{\prime}\right)}{\mathbb{T}\left(x^{\prime}|x\right)}\\
&=&I\left[g\left(\tau,x\right),c\left(x;x^{\prime}\right)\right]+\sigma,
\end{eqnarray}
where we use the shift-invariant property and change of the summation index to have the second equality. 
Finally, by employing the contraction principle, we obtain 
\begin{eqnarray}
\displaystyle \nonumber&&\min_{g\left(\tau,x\right),c\left(x;x^{\prime}\right)}\left\{I\left[g\left(\tau,x\right),c\left(x^{\prime};x\right)\right]|\sum_{x,x^{\prime}\in\Omega}c\left(x,x^{\prime}\right)\log\frac{\mathbb{T}\left(x|x^{\prime}\right)}{\mathbb{T}\left(x^{\prime}|x\right)}=\sigma\right\}\\
\displaystyle \nonumber&&=\min_{g\left(\tau,x\right),c\left(x;x^{\prime}\right)}\left\{I\left[g\left(\tau,x\right),c\left(x;x^{\prime}\right)\right]|\sum_{x,x^{\prime}\in\Omega}c\left(x,x^{\prime}\right)\log\frac{\mathbb{T}\left(x|x^{\prime}\right)}{\mathbb{T}\left(x^{\prime}|x\right)}=\sigma\right\}+\sigma.\\
\end{eqnarray}
Accordingly, we find the GCS: 
\begin{equation}
I\left(-\sigma\right)=I\left(\sigma\right)+\sigma.
\end{equation}

\section{Contraction to Level 2.5 rate function}
The fluctuation of current (flow) plays an essential role to characterize nonequilibrium states. 
For continuous-time Markov jump processes, the explicit form of the joint rate function for the empirical occupation and the empirical jump (reset) 
has been revealed as 
\begin{eqnarray}
\displaystyle \nonumber I\left[\mu\left(x\right),c\left(x;x^{\prime}\right)\right]&=&\displaystyle \sum_{x,x^{\prime}\in\Omega}\left\{\omega\left(x|x^{\prime}\right)\mu\left(x^{\prime}\right)-c\left(x;x^{\prime}\right)\right\}\\
&&+\displaystyle \sum_{x,x^{\prime}\in\Omega}c\left(x;x^{\prime}\right)\log\frac{c\left(x;x^{\prime}\right)}{\omega\left(x|x^{\prime}\right)g\left(x^{\prime}\right)},\label{level2.5}
\end{eqnarray}
where $\omega\left(x|x^{\prime}\right)$ denotes the transition rate of the Markov process from the state $x^{\prime}$ to $x$; 
$\mu\left(x\right)$ and $c\left(x;x^{\prime}\right)$ represent the occupation of the state $x$ and the jump from $x^{\prime}$ to $x$, respectively (also see (\ref{mu1x}) and (\ref{c2xx})). 
The marginal measure $g\left(x\right)$ is given by the shift-invariant property (\ref{shiftccg}). 
This explicit form describes fluctuation of any thermodynamic quantities 
concerned with the current (e.g. heat flow and entropy production) on Markov processes through the contraction principle. 
This rate function (\ref{level2.5}) is known as the {\it Level 2.5} rate function and is derived by various methods. 
A rigorous proof was given by Bertini {\it et al}. \cite{26,27}, 
and a more familiar (heuristic) derivation for physicists was done by Barato and Chetrite by using tilting or spectral technique \cite{28}. 
In this section, under the DTI assumption, we derive an extension of the Level 2.5 rate function to semi-Markov cases. 
Furthermore, from its contraction, we rederive the ordinary Level 2.5 rate function on Markov processes, (\ref{level2.5}). 

\subsection{Level 2.5 rate function for DTI semi-Markov processes}
We introduce an age representation of the rate function (\ref{Igc}), 
which is to express an extension of the Level 2.5 rate function to DTI semi-Markov processes. 
Let us change representation of the semi-Markov process. 
While we described the semi-Markov process by using inter-event interval (waiting time) as $\left(T,X\right)$ up to the previous section, 
in this section, we represent the same process by employing time series for age $A_{t}$, instead of inter-event interval, as $\left(A,X\right)=\left\{A_{t},X_{t}\right\}$ (see Figure 3). 
\begin{figure}[h]
\begin{center}
\includegraphics*[height=5.5cm]{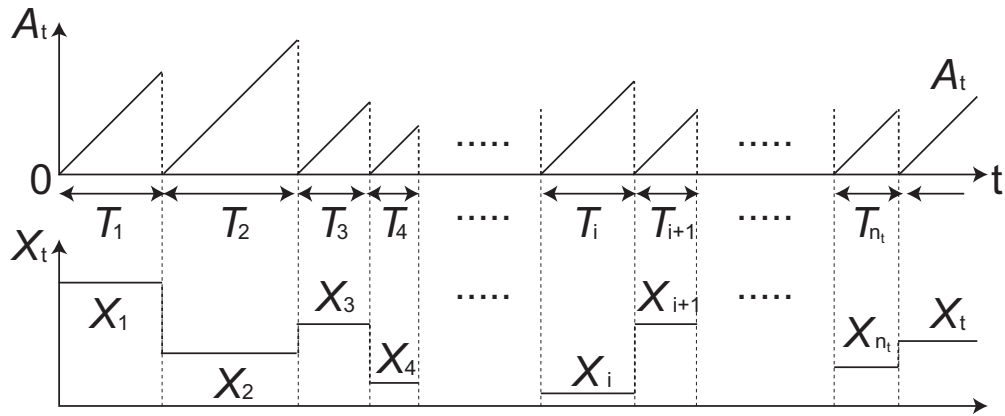}
\caption{The upper graph describes a trajectory of age, $A_{t}$; each pitch of the sawtooth represents inter-event interval $T_{i}$. 
The lower trajectory expresses a time series of state, $X_{t}$.}
\end{center}
\end{figure}
That is, $\left(A,X\right)$ share the same dynamics (probability laws) with $\left(T,X\right)$; the difference between them is only representation. 
Consider an empirical occupation: 
\begin{equation}
\displaystyle \mu_{e}\left(a,x\right):=\frac{1}{t}\int_{0}^{t}dt^{\prime}\,\delta\left(a-A_{t^{\prime}}\right)\delta\left(x-X_{t^{\prime}}\right),
\end{equation}
which represents how often the set $\left(a,x\right)$ appears in the realization $\left(A,X\right)$. 
First, we show a relationship between $\mu_{e}\left(a,x\right)$ and $g_{e}\left(\tau,x\right)$. 
Since the decrease of the occupation $\mu_{e}\left(a,x\right)$ with respect to aging is caused by the occurrence of events at age $a$, we obtain 
\begin{equation}
-\displaystyle \frac{\partial\mu_{e}\left(a,x\right)}{\partial a}=g_{e}\left(a,x\right).\label{diffmug}
\end{equation}
Also, since $\mu_{e}\left(0,x\right)$ expresses the number of jumps (resets) from an arbitrary state to the state $x$ in $\left(A,X\right)$, it can be expressed as 
\begin{equation}
\mu_{e}\left(0,x\right)=g_{e}\left(x\right),\label{mug}
\end{equation}
where $g_{e}\left(x\right)=\Sigma_{x^{\prime}\in\Omega}c_{e}\left(x;x^{\prime}\right)$ represents inflow to the state $x$. 
Solving (\ref{diffmug}) with the boundary condition (\ref{mug}), we have
\begin{equation}
\displaystyle \mu_{e}\left(a,x\right)=\int_{a}^{\infty}d\tau\,g_{e}\left(\tau,x\right),\label{muintg}
\end{equation}
where we use $\mu_{e}\left(0,x\right)=g_{e}\left(x\right)= \int_{0}^{\infty} d\tau\,g_{e}\left(\tau,x\right)$. 
Thus, we find that the correspondence between $\mu_{e}\left(a,x\right)$ and $g_{e}\left(\tau,x\right)$ is a bijection. 
Taking this fact into account, we obtain the rate function for $\mu_{e}\left(a,x\right)$ and $c_{e}\left(x;x^{\prime}\right)$ as 
\begin{eqnarray}
\displaystyle \nonumber I\left[\mu\left(a,x\right),c\left(x;x^{\prime}\right)\right]&=&\displaystyle \sum_{x\in\Omega}\int_{0}^{\infty}da\,\left\{-\frac{\partial\mu\left(a,x\right)}{\partial a}\right\}\log\frac{\left\{-\frac{\partial\mu\left(a,x\right)}{\partial a}\right\}}{\pi\left(\tau|x\right)\mu\left(0,x\right)}\\
&&+\displaystyle \sum_{x,x^{\prime}\in\Omega}c\left(x;x^{\prime}\right)\log\frac{c\left(x;x^{\prime}\right)}{\mathbb{T}\left(x|x^{\prime}\right)\mu\left(0,x\right)},\label{Imuc}
\end{eqnarray}
where we substitute (\ref{diffmug}) and (\ref{mug}) into (\ref{Igc}). 
Note that the following shift-invariant property holds, due to (\ref{mug}): 
\begin{equation}
\displaystyle \mu\left(0,x\right)=\sum_{x^{\prime}\in\Omega}c\left(x^{\prime};x\right)=\sum_{x^{\prime}\in\Omega}c\left(x;x^{\prime}\right).\label{shiftmucc}
\end{equation}
As shown in the next subsection, the contraction of the rate function (\ref{Imuc}) gives Level 2.5 LDP on continuous-time Markov processes 
under the Markov condition (\ref{QMarkov}), i.e., the event rate does not depend on age: $r\left(a;x\right)=r\left(x\right)$. 
Accordingly, we can say that the rate function (\ref{Imuc}) is an extension of the Level 2.5 rate function to DTI semi-Markov processes. 

Finally, we define the occupation distribution of the state $x^{\prime}$ as $\mu_{e}\left(x^{\prime}\right)$, which is a marginal distribution of $\mu_{e}\left(a,x\right)$. 
By using $g_{e}\left(\tau^{\prime},x^{\prime}\right)$, we can represent $\mu_{e}\left(x^{\prime}\right)$ as 
\begin{equation}
\displaystyle \mu_{e}\left(x^{\prime}\right):=\int_{0}^{\infty}da\,\mu_{e}\left(a,x\right)=\int_{0}^{\infty}d\tau^{\prime}\,\tau^{\prime}g_{e}\left(\tau^{\prime},x^{\prime}\right),\label{mu1x}
\end{equation}
where we use (\ref{muintg}). 

\subsection{Contraction to Level 2.5 rate function for Markov processes}
Here, we rederive the Level 2.5 rate function (\ref{level2.5}), by using contraction of the rate function (\ref{Imuc}). 
Consider the rate function on the DTI semi-Markov process, (\ref{Imuc}), with an event rate $r\left(a,x\right)=r\left(x\right)$. 
Then, owing to the contraction principle, Level 2.5 rate function for Markov processes is given by 
\begin{eqnarray}
\displaystyle \nonumber I\left[\mu\left(x\right),c\left(x;x^{\prime}\right)\right]=\min_{\mu\left(a,x\right)}\biggl\{&&I\displaystyle \left[\mu\left(a,x\right),c\left(x;x^{\prime}\right)\right]|\int_{0}^{\infty}da\,\mu\left(a,x\right)=\mu\left(x\right),\\
&&\displaystyle \mu\left(0,x\right)=\sum_{x^{\prime}\in\Omega}c\left(x^{\prime};x\right)=\sum_{x^{\prime}\in\Omega}c\left(x;x^{\prime}\right)\biggr\},\label{mucmin}
\end{eqnarray}
where the second constraint is due to the shift-invariant property (\ref{shiftmucc}). 
Here, we note that the joint occupation $\mu\left(a,x\right)$ is contracted to the state occupation $\mu\left(x\right)$ through (\ref{mu1x}). 
Although we can directly calculate the minimization to obtain the Level 2.5 rate function, we here employ another approach. 
First, we note that the following minimization with respect to $g\left(\tau,x\right)$ instead of $\mu\left(a,x\right)$ is equivalent to that in (\ref{mucmin}): 
\begin{eqnarray}
\displaystyle \nonumber I\left[\mu\left(x\right),c\left(x;x^{\prime}\right)\right]=\min_{g\left(\tau,x\right)}\biggl\{&&I\displaystyle \left[g\left(\tau,x\right),c\left(x;x^{\prime}\right)\right]|\int_{0}^{\infty}d\tau\,\tau g\left(\tau,x\right)=\mu\left(x\right),\\
\displaystyle \nonumber&&\int_{0}^{\infty}d\tau\,g\left(\tau,x\right)=\sum_{x^{\prime}\in\Omega}c\left(x^{\prime};x\right)=\sum_{x^{\prime}\in\Omega}c\left(x;x^{\prime}\right)\biggr\},\\\label{mucming}
\end{eqnarray}
where the rate function $I\left[g\left(\tau,x\right),c\left(x,x^{\prime}\right)\right]$ is given by (\ref{Igc}). 
The first constraint comes from (\ref{mu1x}) and the second one is from (\ref{mug}). 
Hereafter, we consider the minimization in (\ref{mucming}) instead of one in (\ref{mucmin}). 
Next, by substituting the Markov condition $r\left(a,x\right)=r\left(x\right)$ into the explicit form of $I\left[g\left(\tau,x\right),c\left(x;x^{\prime}\right)\right]$, (\ref{Igc}), we have 
\begin{eqnarray}
\displaystyle \nonumber I\left[g\left(\tau,x\right),c\left(x;x^{\prime}\right)\right]&=&\displaystyle \sum_{x\in\Omega}\int_{0}^{\infty}d\tau\,g\left(\tau,x\right)\log g\left(\tau,x\right)+\sum_{x\in\Omega}r\left(x\right)\mu\left(x\right)\\
\displaystyle \nonumber&&-\sum_{x\in\Omega}g\left(x\right)\log r\left(x\right)g\left(x\right)\\
&&+\displaystyle \sum_{x,x^{\prime}\in\Omega}c\left(x;x^{\prime}\right)\log\frac{c\left(x;x^{\prime}\right)}{\mathbb{T}\left(x|x^{\prime}\right)g\left(x^{\prime}\right)},\label{subrinto}
\end{eqnarray}
where $g\left(x\right)$ is given by the shift-invariant property (\ref{shiftccg}). 
Since all terms except the first one are fixed by the constraints, we can simplify the minimization problem in (\ref{mucming}) as 
\begin{eqnarray}
\displaystyle \nonumber\min_{g\left(\tau,x\right)}\biggl\{&&\displaystyle \sum_{x\in\Omega}\int_{0}^{\infty}d\tau\,g\left(\tau,x\right)\log g\left(\tau,x\right)|\int_{0}^{\infty}d\tau\,\tau g\left(\tau,x\right)=\mu\left(x\right),\\
&&\displaystyle \int_{0}^{\infty}d\tau\,g\left(\tau,x\right)=\sum_{x^{\prime}\in\Omega}c\left(x^{\prime};x\right)=\sum_{x^{\prime}\in\Omega}c\left(x;x^{\prime}\right)\biggr\}.
\end{eqnarray}
By employing the Lagrange multiplier method, we find that the function $g^{*}\left(\tau,x\right)$ attaining the above minimization satisfies the following equation: 
\begin{equation}
\displaystyle \sum_{x\in\Omega}\int_{0}^{\infty}d\tau\,g^{*}\left(\tau,x\right)\log g^{*}\left(\tau,x\right)=\sum_{x\in\Omega}g\left(x\right)\log\frac{g^{2}\left(x\right)}{\mu\left(x\right)}-\sum_{x\in\Omega}g\left(x\right).\label{ggent}
\end{equation}
By substituting (\ref{ggent}) into (\ref{subrinto}), we obtain the level 2.5 rate function as 
\begin{eqnarray}
\displaystyle \nonumber I\left[\mu\left(x\right),c\left(x;x^{\prime}\right)\right]&=&\displaystyle \sum_{x\in\Omega}\left\{r\left(x\right)\mu\left(x\right)-g\left(x\right)\right\}\\
&&+\displaystyle \sum_{x,x^{\prime}\in\Omega}c\left(x;x^{\prime}\right)\log\frac{c\left(x;x^{\prime}\right)}{\mathbb{T}\left(x|x^{\prime}\right)r\left(x^{\prime}\right)g\left(x^{\prime}\right)},\label{I2.5cont}
\end{eqnarray}
where we use the shift-invariant property (\ref{shiftccg}). 
Finally, recalling that the transition rate of Markov processes can be represented as $\omega\left(x|x^{\prime}\right)=\mathbb{T}\left(x|x^{\prime}\right)r\left(x^{\prime}\right)$ (i.e., $\Sigma_{x^{\prime}\in\Omega}\omega\left(x^{\prime}|x\right)\mu\left(x\right)=r\left(x\right)\mu\left(x\right)$), 
and $g\left(x\right)=\Sigma_{x^{\prime}\in\Omega}c\left(x^{\prime};x\right)$, we find that the explicit form (\ref{I2.5cont}) is equivalent to (\ref{level2.5}). 

\section{Summary}
We have derived the explicit form of the rate function for semi-Markov processes with respect to the empirical triplet (\ref{je3}). 
Also, we have shown that the explicit form can be decomposed into point-process and Markov-process parts under the DTI assumption. 
In addition, by exploiting the contraction principle to the decomposed rate function, we have elucidated that the FT (Gallavotti-Cohen Symmetry) holds for DTI semi-Markov cases. 
Furthermore, we have found that the age representation of our rate function for semi-Markov processes gives an extension version of the Level 2.5 rate function for Markov processes. 

The explicit forms obtained in this paper contribute to analysis for the age-structured population dynamics. 
We will show an application of our rate function to biological problems in our next paper \cite{33}. 

\section*{Acknowledgments}
We thank Rosemary J. Harris and Massimo Gavallaro for fruitful discussion. 
This research is supported by JSPS KAKENHI Grant Number JP16K17763, JP16H06155 and JST PRESTO Grant Number JPMJPR15E4, Japan. 

\appendix
\section{Sanov's theorem for Markov processes}
Here, we show a brief derivation of the rate function (\ref{IJ4}); A rigorous proof is shown in references \cite{01,02,04}. 
For simplicity of calculation, we deal with 1-dimensional Markov processes; an extension to multidimensional processes is straightforward. 
Consider a time-discrete Markov process $\left(X\right)=\left\{X_{i}\right\}$ with an ergodic transition probability $\mathbb{T}\left(x|x^{\prime}\right)$, 
and its scaled cumulant generating function \cite{04} for the pair empirical measure $J_{e}\left(x,x^{\prime}\right):=\left(1/n\right)\Sigma_{i=1}^{n}\delta_{x,X_{i+1}}\delta_{x^{\prime},X_{i}}$, 
where we again assume the periodic condition $X_{i+1}=X_{1}$. 
Then, we have
\begin{eqnarray}
\displaystyle \nonumber\lambda\left[k\right]&:=&\displaystyle \lim_{n\rightarrow\infty}\frac{1}{n}\log\left\langle\exp\left\{n\sum_{x,x^{\prime}\in\Omega}k\left(x,x^{\prime}\right)J_{e}\left(x,x^{\prime}\right)\right\}\right\rangle\\
\displaystyle \nonumber&=&\displaystyle \lim_{n\rightarrow\infty}\frac{1}{n}\log\left\langle e^{\sum_{i=1}^{n}k\left(X_{i+1},X_{i}\right)}\right\rangle\\
&=&\displaystyle \lim_{n\rightarrow\infty}\frac{1}{n}\log\sum_{\left\{x_{i}\right\}\in\Omega^{n}}e^{k\left(x_{0},x_{n}\right)}\prod_{i=1}^{n-1}e^{k\left(x_{i+1},x_{i}\right)}\mathbb{T}\left(x_{i+1}|x_{i}\right)\rho\left(x_{1}\right),
\end{eqnarray}
where $\left\langle\cdot\right\rangle$ represents the average over all paths $\left\{X_{i}|1\leq i\leq n\right\}$ with a path probability $\Pi_{i=1}^{n-1}\mathbb{T}\left(x_{i+1}|x_{i}\right)\rho\left(x_{1}\right)$; $\rho\left(\cdot\right)$ is an arbitrary initial distribution. 
By using diagonalization of $e^{k\left(\cdot,\cdot^{\prime}\right)}\mathbb{T}\left(\cdot|\cdot^{\prime}\right)$ and taking the limit $ n\rightarrow\infty$ into account, 
we find that the scaled cumulant generating function $\lambda\left[k\right]$ is given by logarithm of the largest eigenvalue of $e^{k\left(\cdot,\cdot^{\prime}\right)}\mathbb{T}\left(\cdot|\cdot^{\prime}\right)$. 
That is, by employing the corresponding right eigenvector $v_{k}\left(\cdot\right)$ (the right eigenvector corresponding to the largest eigenvalue), we have 
\begin{equation}
\displaystyle \sum_{y\in\Omega}e^{k\left(x,y\right)}\mathbb{T}\left(x|y\right)v_{k}\left(y\right)=e^{\lambda\left[k\right]}v_{k}\left(x\right),\label{a2}
\end{equation} 
where the uniqueness of the largest eigenvalue and the positivity of the corresponding eigenvector are guaranteed by the Perron-Frobenius theorem. 
From the G\"{a}rtner-Ellis theorem \cite{04}, the Legendre transform of $\lambda\left[k\right]$ gives the rate function for the pair empirical measure $J_{e}\left(x,y\right)$: 
\begin{equation}
I\displaystyle \left[J\right]=\max_{k}\left\{\sum_{x,y\in\Omega}J\left(x,y\right)k\left(x,y\right)-\lambda\left[k\right]\right\}.\label{a1}
\end{equation}
To solve the maximization, we calculate 
\begin{equation}
k^{*}\displaystyle \left(\cdot,\cdot^{\prime}\right):=\arg\max_{k}\left\{\sum_{x,y\in\Omega}J\left(x,y\right)k\left(x,y\right)-\lambda\left[k\right]\right\}.
\end{equation}
From the variation of (\ref{a1}) with respect to $k\left(x,y\right),\ k^{*}\left(x,y\right)$ satisfies 
\begin{equation}
\left.\frac{\delta\lambda\left[k\right]}{\delta k\left(x,y\right)}\right|_{k=k*}=J\left(x,y\right).\label{a5}
\end{equation}
To compute the variation of the left hand side, we consider a procedure like perturbation methods. 
Now, analyze the following perturbed equation: 
\begin{eqnarray}
\displaystyle \nonumber\sum_{y\in\Omega}e^{k\left(x,y\right)+\delta k\left(x,y\right)}\mathbb{T}\left(x|y\right)\left\{v_{k}\left(y\right)+\delta v_{k}\left(y\right)\right\}=e^{\lambda\left[k\right]+\delta\lambda\left[k\right]}\left\{v_{k}\left(x\right)+\delta v_{k}\left(x\right)\right\}.\\
\end{eqnarray}
Evaluation within the first order of $\delta$ leads to 
\begin{eqnarray}
\displaystyle \nonumber\sum_{y\in\Omega}e^{k\left(x,y\right)}\mathbb{T}\left(x|y\right)\delta v_{k}\left(y\right)+\sum_{y\in\Omega}e^{k\left(x,y\right)}\mathbb{T}\left(x|y\right)v_{k}\left(y\right)\delta k\left(x,y\right)\\
=e^{\lambda\left[k\right]}\delta v_{k}\left(x\right)+e^{\lambda\left[k\right]}v_{k}\left(x\right)\delta\lambda\left[k\right],\label{a3}
\end{eqnarray}
where we use (\ref{a2}) to simplify the equation. 
Furthermore, by applying the corresponding left eigenvector $u_{k}\left(\cdot\right)$ to both sides of (\ref{a3}) from the left side, we have 
\begin{eqnarray}
\displaystyle \nonumber\sum_{x,y\in\Omega}u_{k}\left(x\right)e^{k\left(x,y\right)}\mathbb{T}\left(x|y\right)\delta v_{k}\left(y\right)+\sum_{x,y\in\Omega}u_{k}\left(x\right)e^{k\left(x,y\right)}\mathbb{T}\left(x|y\right)v_{k}\left(y\right)\delta k\left(x,y\right)\\
=e^{\lambda\left[k\right]}\displaystyle \sum_{x\in\Omega}u_{k}\left(x\right)\delta v_{k}\left(x\right)+e^{\lambda\left[k\right]}\sum_{x\in\Omega}u_{k}\left(x\right)v_{k}\left(x\right)\delta\lambda\left[k\right].\label{a4}
\end{eqnarray}
Taking into account the fact that $u_{k}\left(\cdot\right)$ represents the left eigenvector of $e^{k\left(\cdot,\cdot^{\prime}\right)}\mathbb{T}\left(\cdot|\cdot^{\prime}\right)$:
\begin{equation}
\displaystyle \sum_{x\in\Omega}u_{k}\left(x\right)e^{k\left(x,y\right)}\mathbb{T}\left(x|y\right)=e^{\lambda\left[k\right]}u_{k}\left(y\right),\label{a6}
\end{equation}
we can cancel the first terms in both sides of (\ref{a4}). 
Then, after simplifying (\ref{a4}), we obtain 
\begin{equation}
\displaystyle \frac{\delta\lambda\left[k\right]}{\delta k\left(x,y\right)}=\frac{u_{k}\left(x\right)e^{k\left(x,y\right)-\lambda\left[k\right]}\mathbb{T}\left(x|y\right)v_{k}\left(y\right)}{\sum_{z\in\Omega}u_{k}\left(z\right)v_{k}\left(z\right)}.
\end{equation}
Accordingly, from (\ref{a5}), $k^{*}\left(x,y\right)$ satisfies 
\begin{equation}
J\displaystyle \left(x,y\right)=\frac{u_{k^{*}}\left(x\right)e^{k^{*}\left(x,y\right)-\lambda\left[k^{*}\right]}\mathbb{T}\left(x|y\right)v_{k^{*}}\left(y\right)}{\sum_{z\in\Omega}u_{k^{*}}\left(z\right)v_{k^{*}}\left(z\right)}.\label{a7}
\end{equation}
Also, we define a marginal distribution $G\left(\cdot\right)$ as 
\begin{equation}
G\displaystyle \left(y\right):=\sum_{x\in\Omega}J\left(x,y\right)=\sum_{x\in\Omega}J\left(y,x\right)=\frac{u_{k^{*}}\left(y\right)v_{k^{*}}\left(y\right)}{\sum_{z\in\Omega}u_{k^{*}}\left(z\right)v_{k^{*}}\left(z\right)},\label{a10}
\end{equation}
where we use (\ref{a6}). 
By solving (\ref{a7}) with respect to $e^{k^{*}\left(x,y\right)-\lambda\left[k^{*}\right]}$, we have 
\begin{equation}
e^{k^{*}\left(x,y\right)-\lambda\left[k^{*}\right]}=\displaystyle \frac{J\left(x,y\right)\sum_{z\in\Omega}u_{k^{*}}\left(z\right)v_{k^{*}}\left(z\right)}{u_{k^{*}}\left(x\right)\mathbb{T}\left(x|y\right)v_{k^{*}}\left(y\right)}.\label{a8}
\end{equation}
Noting (\ref{a1}), we can express the rate function by using $k^{*}\left(\cdot,\cdot^{\prime}\right)$ as 
\begin{equation}
I\displaystyle \left[J\right]=\sum_{x,y\in\Omega}J\left(x,y\right)\log e^{k^{*}\left(x,y\right)-\lambda\left[k^{*}\right]}.\label{a9}
\end{equation}
By substituting (\ref{a8}) into (\ref{a9}), we get 
\begin{eqnarray}
\displaystyle \nonumber I\left[J\right]=\sum_{x,y\in\Omega}J\left(x,y\right)\log\frac{J\left(x,y\right)}{\mathbb{T}\left(x|y\right)}+\sum_{x,y\in\Omega}J\left(x,y\right)\log\frac{\sum_{z\in\Omega}u_{k^{*}}\left(z\right)v_{k^{*}}\left(z\right)}{u_{k^{*}}\left(x\right)v_{k^{*}}\left(y\right)}.\\\label{a11}
\end{eqnarray}
Finally, by using the shift-invariant property (\ref{a10}), we rewrite the second term in (\ref{a11}) as 
\begin{eqnarray}
\displaystyle \nonumber\sum_{x,y\in\Omega}J\left(x,y\right)\log\frac{\sum_{z\in\Omega}u_{k^{*}}\left(z\right)v_{k^{*}}\left(z\right)}{u_{k^{*}}\left(x\right)v_{k^{*}}\left(y\right)}\\
=\displaystyle \sum_{x,y\in\Omega}J\left(x,y\right)\log\frac{\sum_{z\in\Omega}u_{k^{*}}\left(z\right)v_{k^{*}}\left(z\right)}{u_{k^{*}}\left(y\right)v_{k^{*}}\left(y\right)}=\sum_{x,y\in\Omega}J\left(x,y\right)\log\frac{1}{G\left(y\right)},
\end{eqnarray}
where we use the expression of $G\left(\cdot\right)$, (\ref{a10}). 
Thus, we obtain the explicit form of the rate function as 
\begin{equation}
I\displaystyle \left[J\right]=\sum_{x,y\in\Omega}J\left(x,y\right)\log\frac{J\left(x,y\right)}{\mathbb{T}\left(x|y\right)G\left(y\right)}.\label{aadd1}
\end{equation}
By extending this calculation to 2-dimensional cases, we can find (\ref{IJ4}). 

\section{LDP on point processes}
In this appendix, we introduce point processes \cite{39} and their LDP. 
Suppose an inter-event time interval sequence $\left(T\right):=\left\{T_{i}\right\}$, where each element $T_{i}$ is distributed with a probability density, 
\begin{equation}
\pi\left(\tau\right):=r\left(\tau\right)e^{-\int_{0}^{\tau}r\left(a\right)da}.\label{pointpi}
\end{equation}
Here, $r\left(a\right)$ represents the event rate, i.e., the probability that an event occurs at age $a$. 
The age means the elapsed time after the previous event occurs. 
Thus, we regard $r\left(a\right)$ as a simple version of $r\left(a,x\right)$ introduced in section 2. 
The process $\left(T\right)$ generated by (\ref{pointpi}) is a kind of point processes. 
If $r\left(a\right)$ is constant, the point process is reduced to a homogeneous Poisson point process. 

For the above process, we consider the rate function of the following ``time-normalized" empirical measure: 
\begin{equation}
g_{e}\displaystyle \left(\tau\right):=\frac{1}{t}\sum_{i=1}^{n_{t}}\delta\left(\tau-T_{i}\right),\label{b1}
\end{equation}
which measures how often the inter-event interval $\tau$ appears in the sequence $\left(T\right)$. 
Note that this measure is normalized as $\int_{0}^{\infty} d\tau\,\tau g_{e}\left(\tau\right)=1$ at $ t\rightarrow\infty$. 
According to the procedure in section 3, to calculate the rate function, we firstly consider the ``number-normalized" empirical measure: 
\begin{equation}
G_{e}\displaystyle \left(\tau\right):=\frac{1}{n}\sum_{i=1}^{n}\delta\left(\tau-T_{i}\right),
\end{equation}
which is normalized as $\int_{0}^{\infty} d\tau\,G_{e}\left(\tau\right)=1$, differently from (\ref{b1}). 
Now, since each element $T_{i}$ is independent and identically distributed (IID) from $\pi\left(\cdot\right)$, 
Sanov's theorem for IID \cite{04} leads the explicit form of the rate function for $G_{e}\left(\tau\right)$ as 
\begin{equation}
\displaystyle \tilde{I}\left[G\left(\tau\right)\right]=\int_{0}^{\infty}d\tau\,G\left(\tau\right)\log\frac{G\left(\tau\right)}{\pi\left(\tau\right)}.\label{b2}
\end{equation}
By using the random time change trick shown in subsection 3.2, we obtain 
\begin{equation}
I\left[g\left(\tau\right)\right]=\left\{\int_{0}^{\infty}d\tau\,g\left(\tau\right)\right\}\tilde{I}\left[\frac{g\left(\tau\right)}{\int_{0}^{\infty}d\tau\,g\left(\tau\right)}\right],\label{b3}
\end{equation}
where $I\left[g\left(\tau\right)\right]$ is the rate function of $g_{e}\left(\tau\right)$. 
Substituting (\ref{b2}) into (\ref{b3}), we finally find the explicit form:
\begin{equation}
I\displaystyle \left[g\left(\tau\right)\right]=\int_{0}^{\infty}d\tau\,g\left(\tau\right)\log\frac{g\left(\tau\right)}{\pi\left(\tau\right)g},
\end{equation}
where the constant $g$ is $g:= \int_{0}^{\infty} d\tau\,g\left(\tau\right)$, which represents the number of events per unit time, that is $n_{t}/t$.

\end{document}